\documentclass[12pt,preprint]{aastex}
\usepackage{gensymb} 
\usepackage[version=4]{mhchem}
\usepackage{amsmath}
\usepackage[colorlinks=true,linkcolor=blue,citecolor=blue]{hyperref}
\shorttitle{Head Tail Radio Sources}
\shortauthors{Pal and Kumari}

\begin{document}

\title{Head Tail Radio Sources from VLA FIRST Survey }

\author{Sabyasachi Pal}

\affil{Indian Centre for Space Physics, 43 Chalantika, Garia Station Road, 700084, India\\
Midnapore City College, Kuturia, Bhadutala, West Bengal, 721129, India}

\and

\author{Shobha Kumari}

\affil{Midnapore City College, Kuturia, Bhadutala, West Bengal, 721129, India}

\begin{abstract}
The study of Head Tail (HT) radio galaxies track the information of associated galaxy clusters. With the help of the VLA FIRST survey at 1.4 GHz, we detected 607 new HT radio sources, among them, 398 are Wide Angle Tail (WAT) and 216 are Narrow-Angle Tail (NAT) sources. NAT sources generally have `V' shaped structure with an opening angle less than ninety degrees and for WAT sources opening angle between the jets is more than ninety degrees. We found that almost 80 per cent of our sources are associated with a known galaxy cluster. We mentioned various useful physical properties of these HT sources. Taking advantage of a large sample of newly discovered HT sources, various statistical studies have been done. The luminosity range of sources presented in the current paper is $10^{39}$ $\leq$ $L_{1.4GHz}$ $\leq$ $10^{43}$ erg sec$^{-1}$. We identified optical counterparts for 193 WAT and 104 NAT sources. The sources are found up to redshift 2.08.
\end{abstract}


\section{Introduction}
\label{sec:intro}
The radio jets are the signature of the outflow of energy from the central active galactic nuclei (AGN).
Head Tail (HT) radio galaxies are classified based on alignment and the structure of the radio jets with respect to the central host galaxy.
The jets of HT galaxies are bent into {\it C}, {\it V} or {\it L} like shapes \citep{Ry68, Ru76, Bl00, Pr11, De14}.
Based on the degree of bending of two opposite jets, these types of galaxies have first classified \citep{Ow76} into two categories, namely the Wide Angle Tailed (WAT) radio sources, and Narrow-Angle Tailed (NAT) radio sources respectively. 

HT radio galaxies are generally found in the rich cluster of galaxies \citep{Bu90}. So one can use HT galaxies as a tracer of galaxy cluster which is the building block of our Universe. The structure and shape of HT radio galaxies depend on the angle and alignment of two opposite radio jets with their common core. Using an automated method, \citet{Pr11} presented a catalogue of sources including bent sources from the VLA FIRST survey.
\citet{Ry68} named these sources as HT sources. \citep{Ow76} classified these sources as Wide Angle Tail (WAT) and Narrow-Angle Tail (NAT) based on the bending angle of two opposite jets. 
3C 465 is the best example of WAT class radio sources \citep{Ei84, Ei02, Ha05} and NGC 1265 \citep{Ry68, O'D86} is the best example of NAT sources.
The `ram-pressure model was first proposed by \citep{Be79} to explain the bending of jets in HT sources and later the model is used by many authors \citep{Ba85a, Va81}.
When the material density of the radio jets is less than the density of the surrounding medium, the buoyancy force comes into action.
It pushes the lobes to the regions of ICM where the density of the jet is equal to that of the surrounding medium \citep{Gu73, Sa96} as result jets are bends.
Investigating the WAT sources found in the Abell cluster region \citet{Sa00} hints that the merger of the galaxies may also be a reason for these kinds of sources.
Based on the radio luminosity and its distribution on the radio map, the radio sources are categorized into two classes, namely FR-I and FR-II \citep{Fa74a}.
For HT sources, it is found that most of the WATs and NATs are edge dimmed sources; i.e., FR-I type of sources \citep{Go00}.

We made a systematic search to create a complete list of HT sources from the high-resolution VLA FIRST survey. This paper is arranged in the respective sequence: in Section 2, we sequentially describe our search procedure, which includes the FIRST survey, the search strategy, defining the WATs and NATs, and the respective optical counterpart of each source.
The next section (Section 3) is dedicated to the result.
The Result section deals with the source catalogue, radio properties, like radio luminosity, spectral index, and the physical parameters e.g., angle of the radio sources.
In the last section, (Section 4) we discuss our whole work and the outcomes.

Following cosmology parameters we have used in this paper : $H_0 = 67.4$ km s$^{-1}$ Mpc$^{-1}$, $\Omega_m = 0.315$ and $\Omega_{vac} = 0.685$ \citep{Ag20}.
The spectral index ($\alpha$) is defined by $S_{\nu} \propto \nu^\alpha$ where $S_{\nu}$ is the flux density at frequency $\nu$.

\begin{figure*}
	\vbox{
		\centerline{
			\includegraphics[height=4.2cm,width=4.2cm]{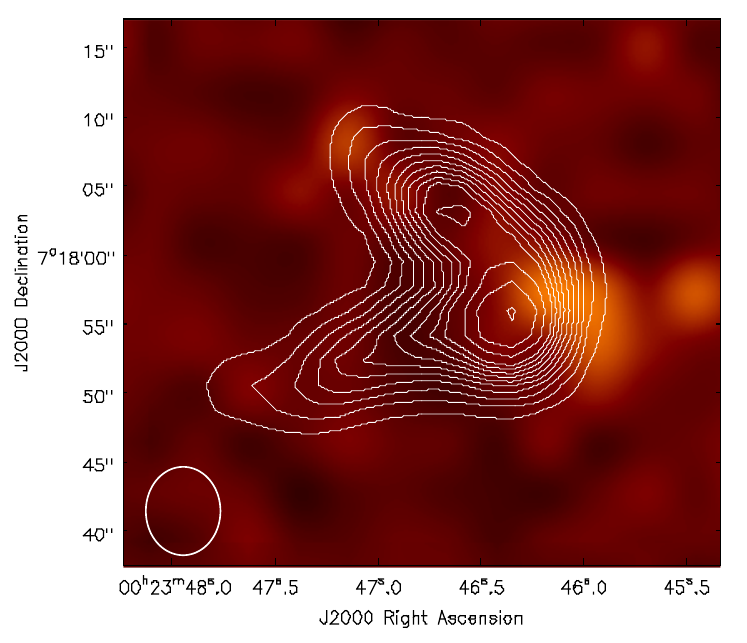}
			\includegraphics[height=4.2cm,width=4.2cm]{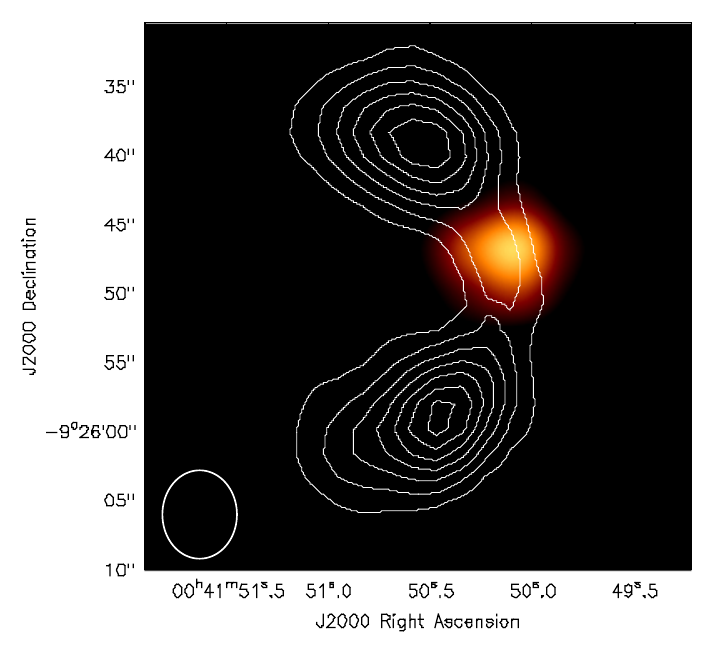}
			\includegraphics[height=4.2cm,width=4.2cm]{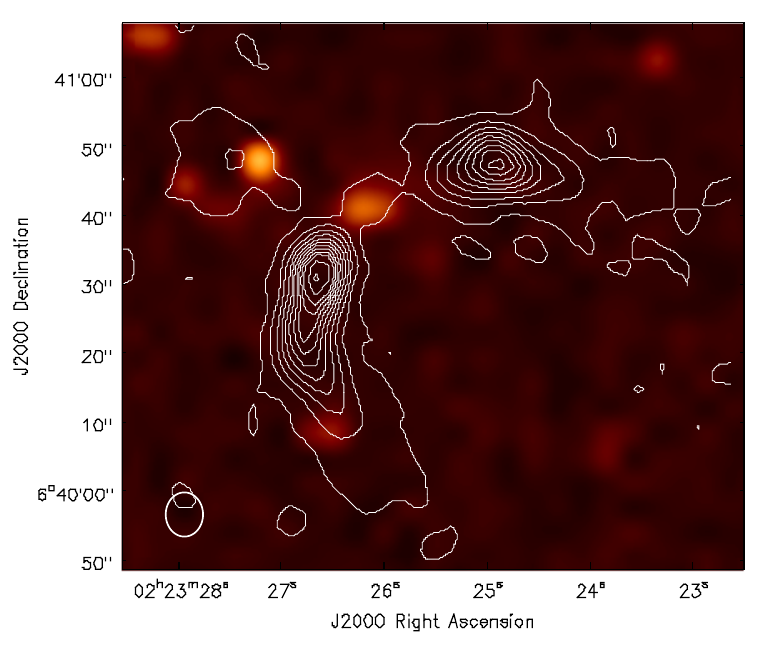}
		}
	}
	\vbox{
		\centerline{
			\includegraphics[height=4.2cm,width=4.2cm]{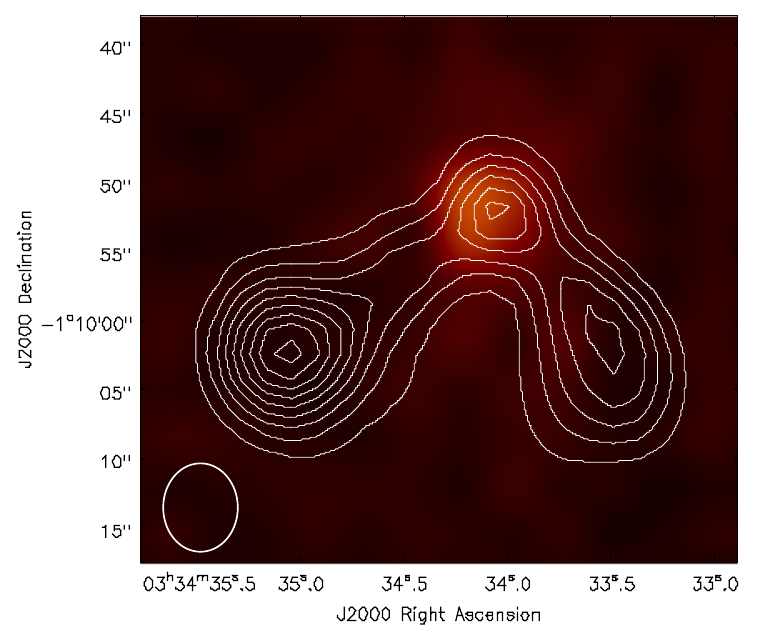}	
			\includegraphics[height=4.2cm,width=4.2cm]{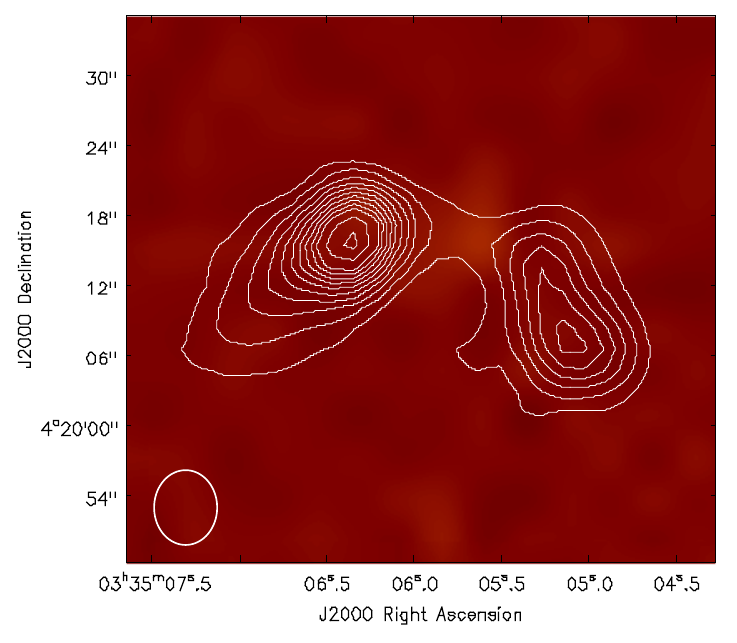}
		        \includegraphics[height=4.2cm,width=4.2cm]{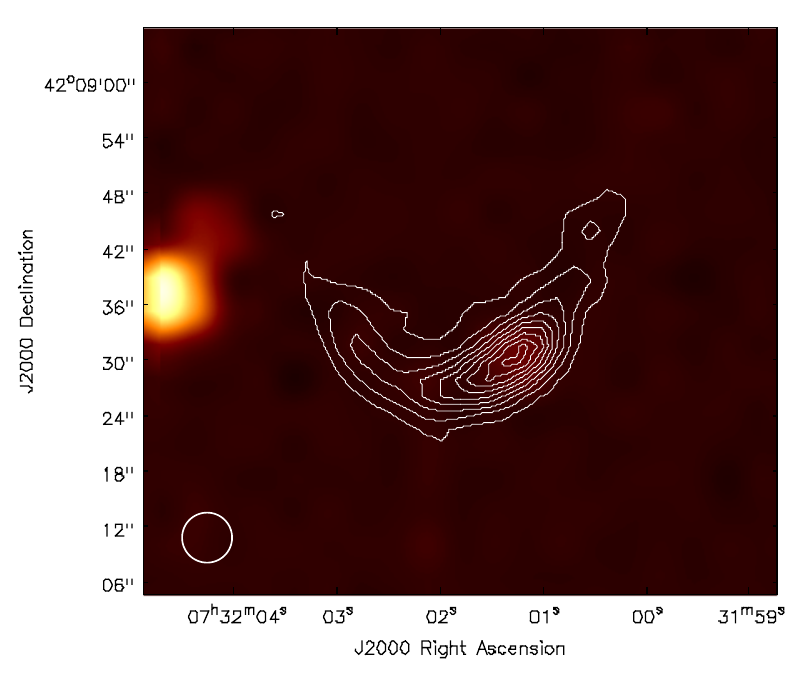}
		}
	}
	\vbox{
	\centerline{
		\includegraphics[height=4.2cm,width=4.2cm]{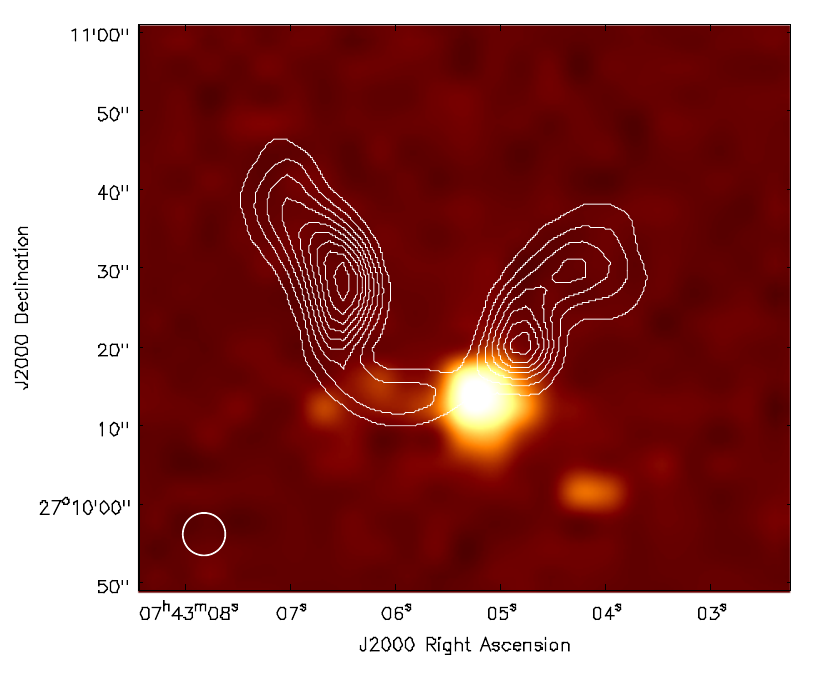}
		\includegraphics[height=4.2cm,width=4.2cm]{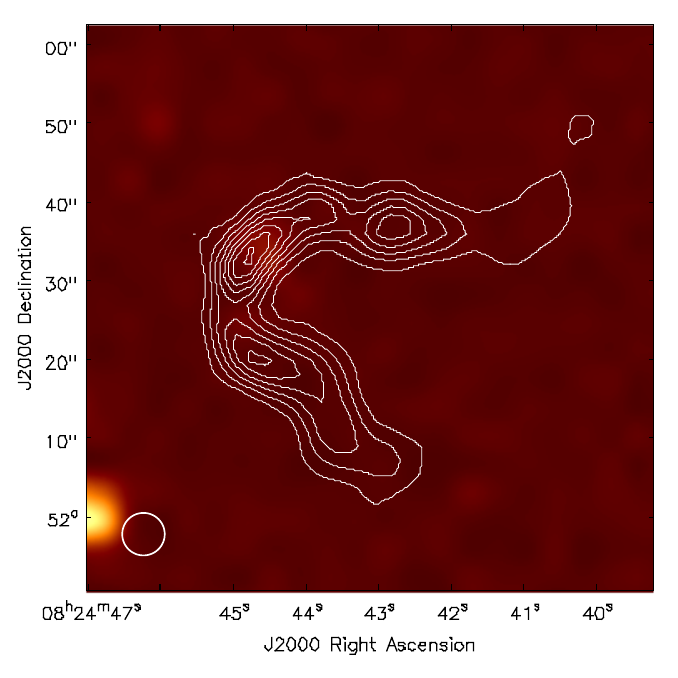}
		\includegraphics[height=4.2cm,width=4.2cm]{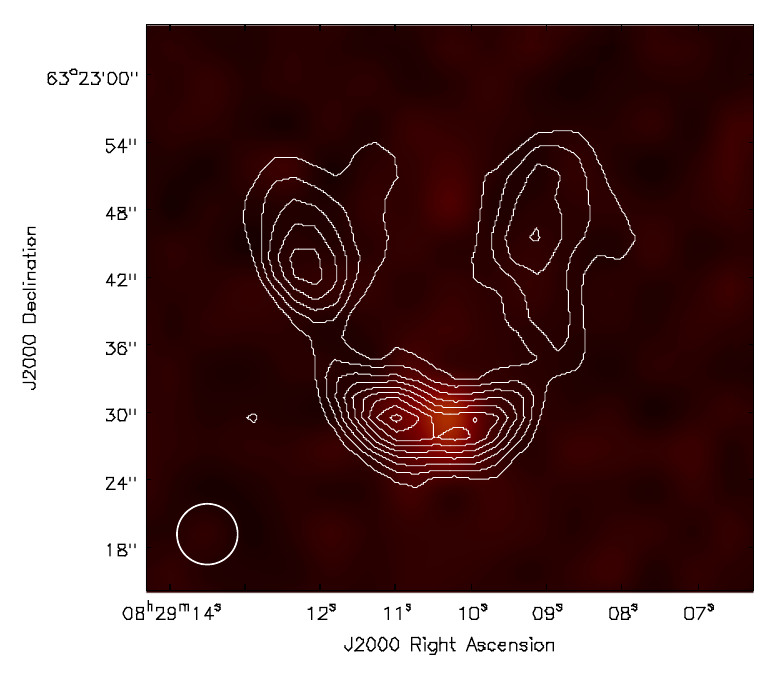}
	}
	}

	\caption{Examples of nine Narrow-Angle Tail (NAT) radio sources (contours) overlaid on the DSS2 red image (greyscale) using the FIRST survey at 1400 MHz.}
\end{figure*}

\begin{figure*}
	\vbox{
		\centerline{
			\includegraphics[height=4.2cm,width=4.2cm]{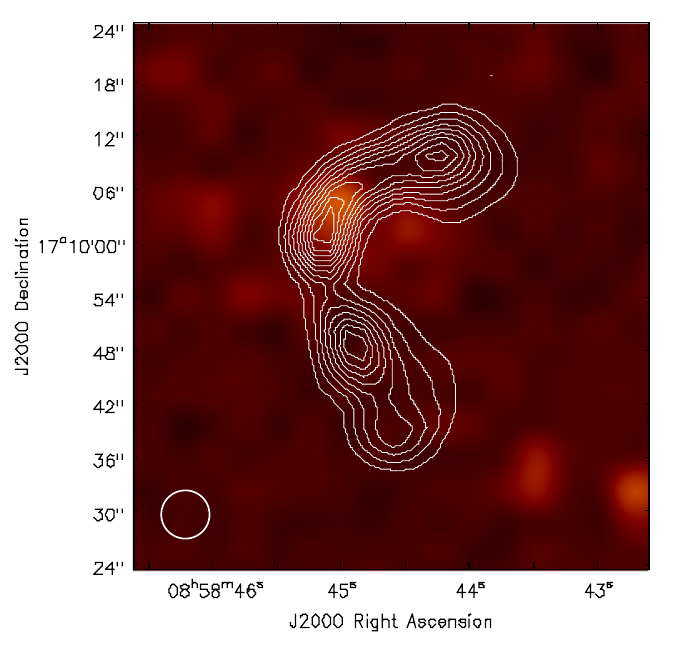}
			\includegraphics[height=4.2cm,width=4.2cm]{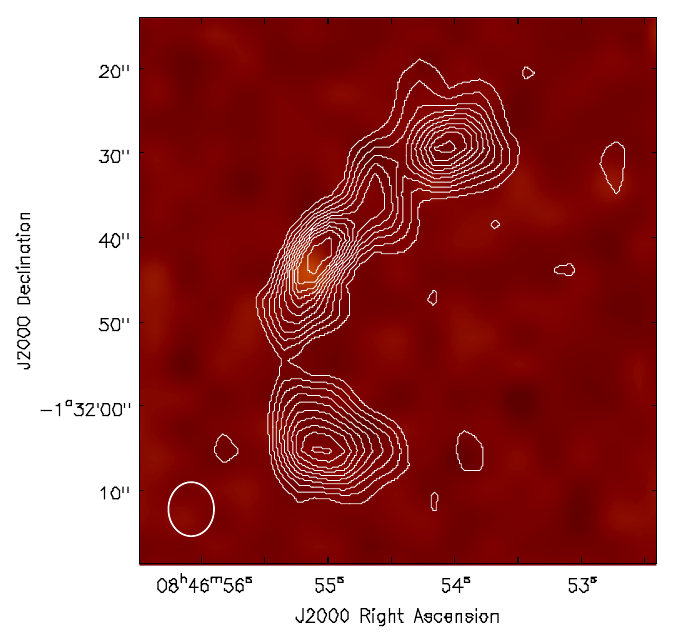}
			\includegraphics[height=4.2cm,width=4.2cm]{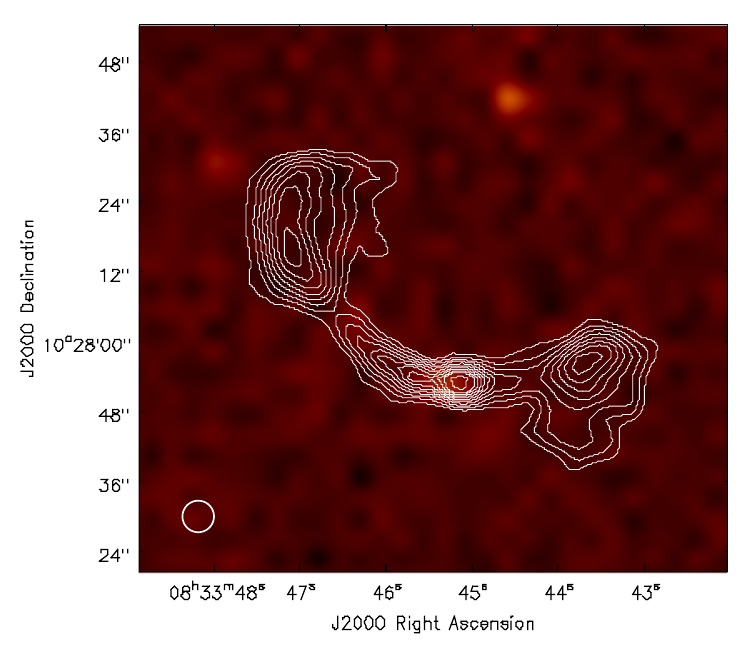}
		}
	}
	\vbox{
		\centerline{	
			\includegraphics[height=4.2cm,width=4.2cm]{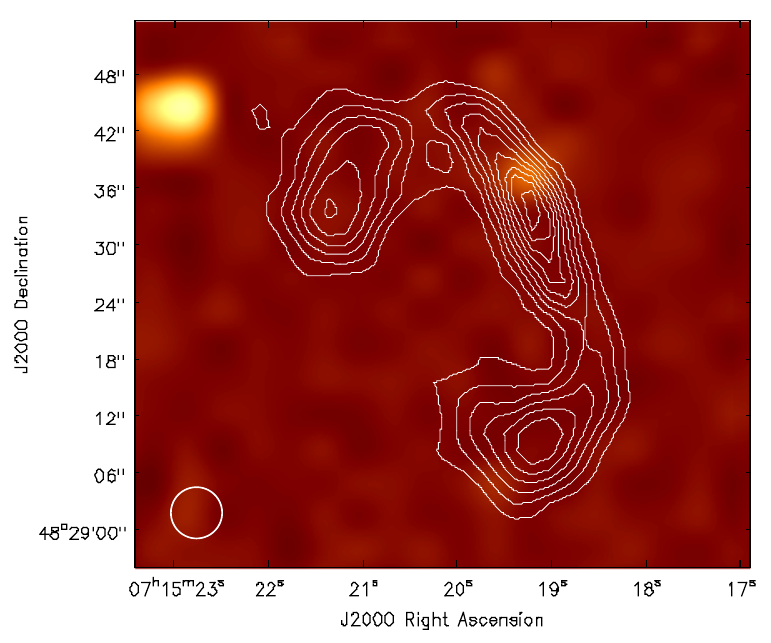}
			\includegraphics[height=4.2cm,width=4.2cm]{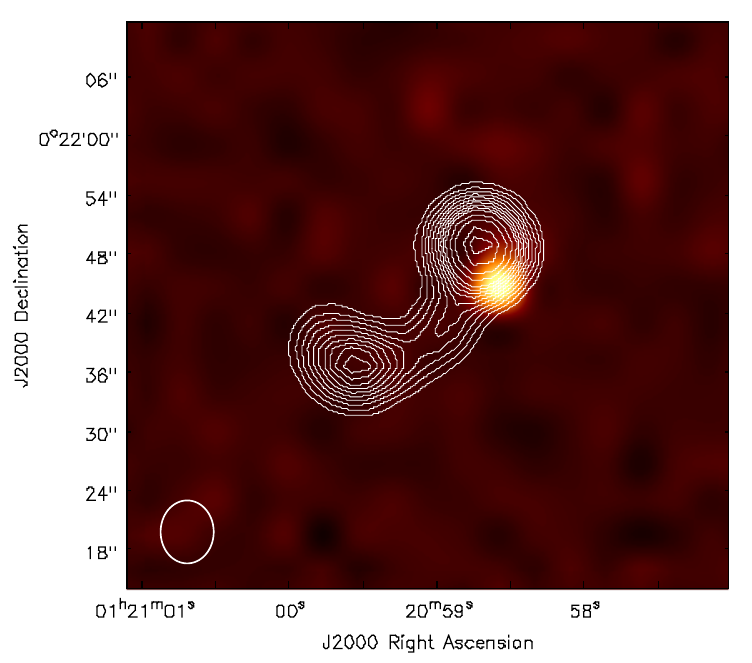}
			\includegraphics[height=4.2cm,width=4.2cm]{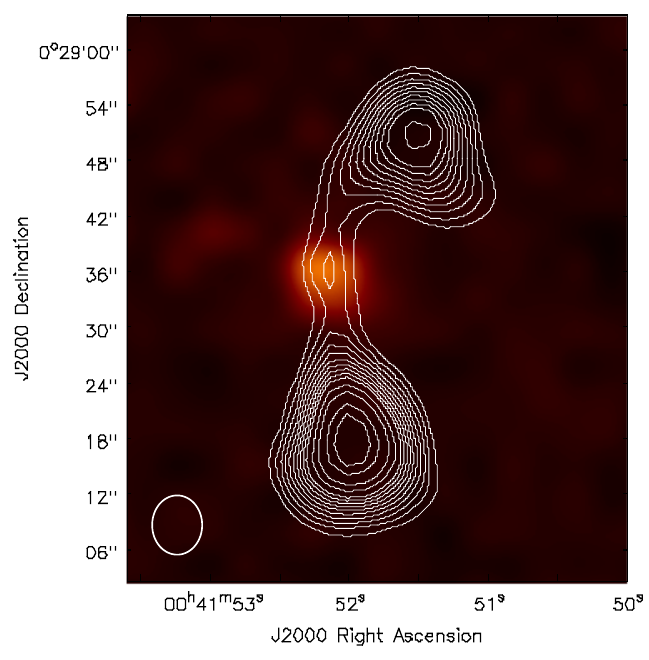}
		}
	}
	\vbox{
		\centerline{
			\includegraphics[height=4.2cm,width=4.2cm]{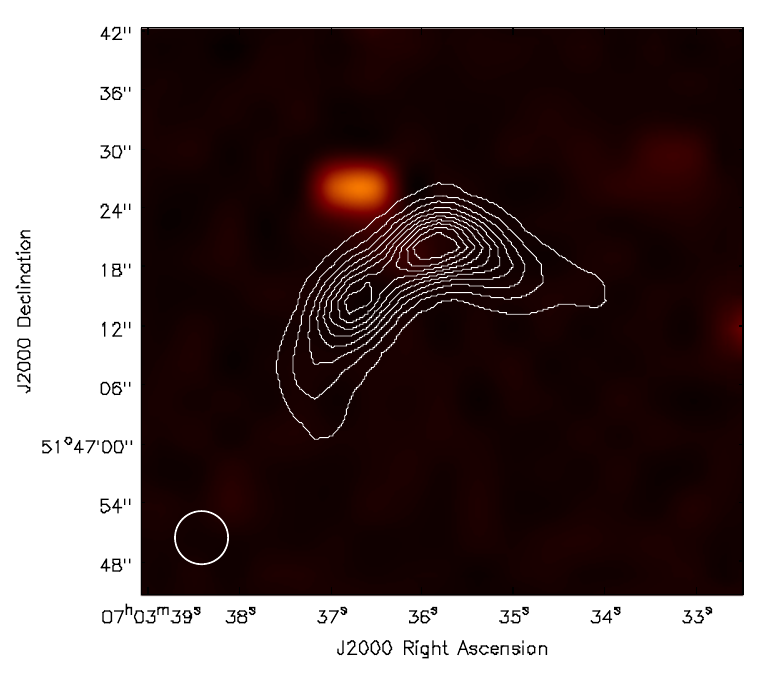}
			\includegraphics[height=4.2cm,width=4.2cm]{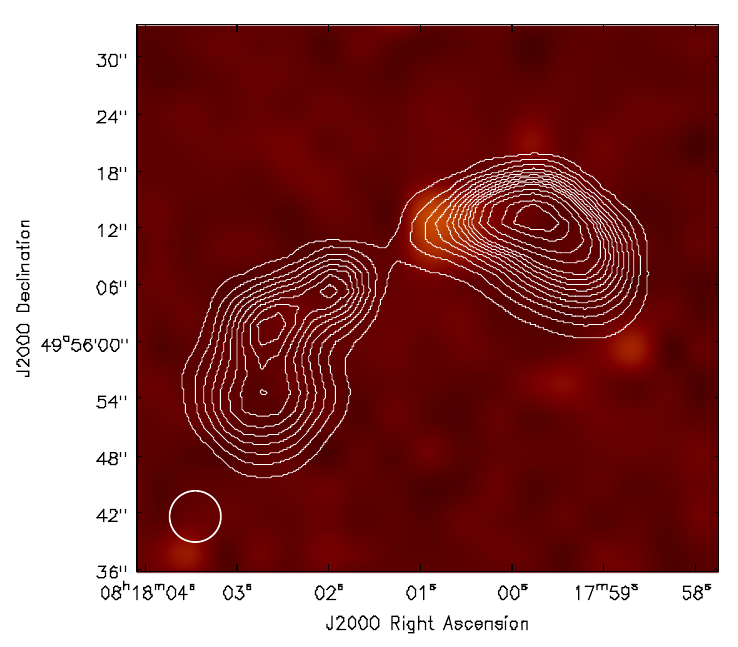}
			\includegraphics[height=4.2cm,width=4.2cm]{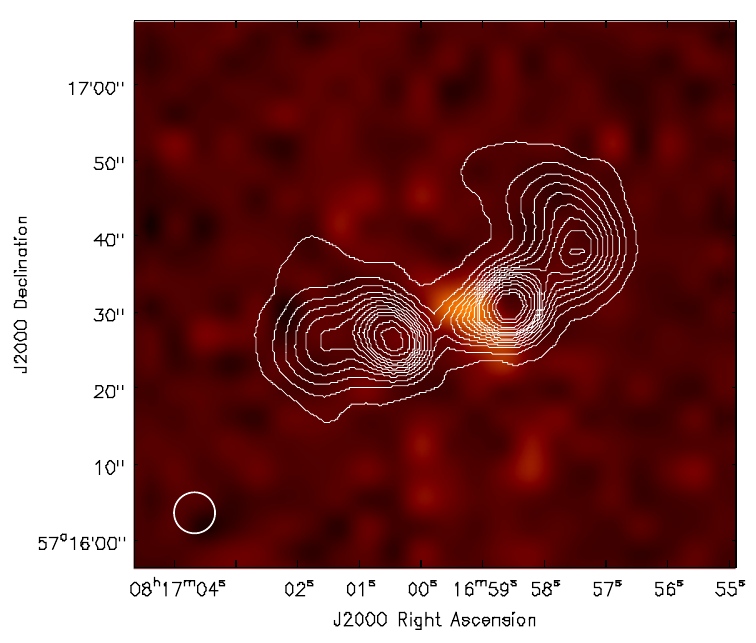}
		}
	}
	        \vbox{
		\centerline{
		\includegraphics[height=4.2cm,width=4.2cm]{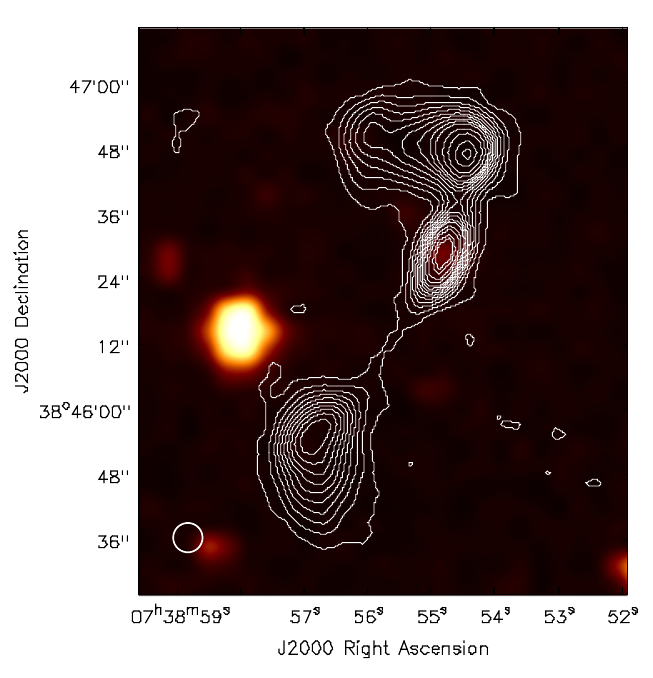}
		\includegraphics[height=4.2cm,width=4.2cm]{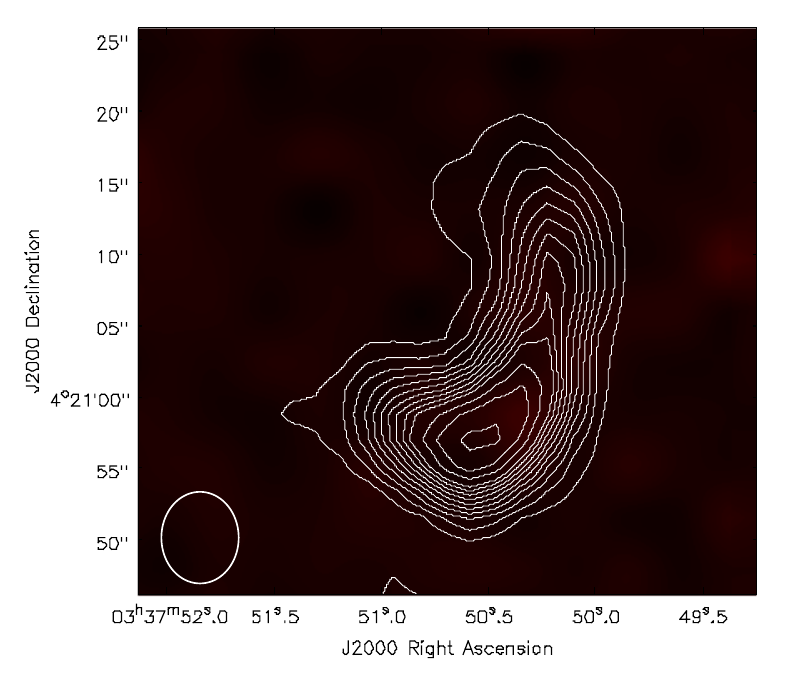}
		\includegraphics[height=4.2cm,width=4.2cm]{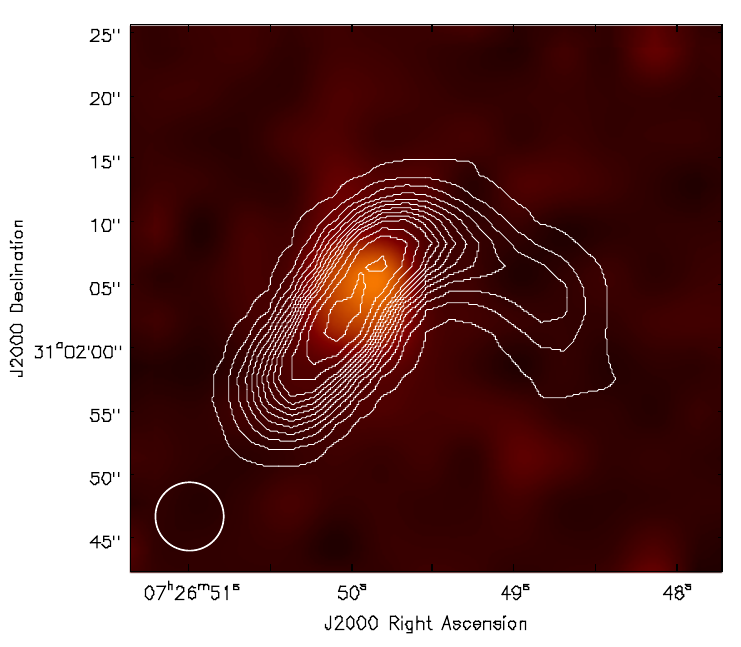}	
		}
	}

	\caption{Examples of nine Narrow-Angle Tail (NAT) radio sources (contours) overlaid on the DSS2 red image (greyscale) using the FIRST survey at 1400 MHz.}
\end{figure*}

\begin{figure*}
	\centering
\includegraphics[width=8.1cm,origin=c]{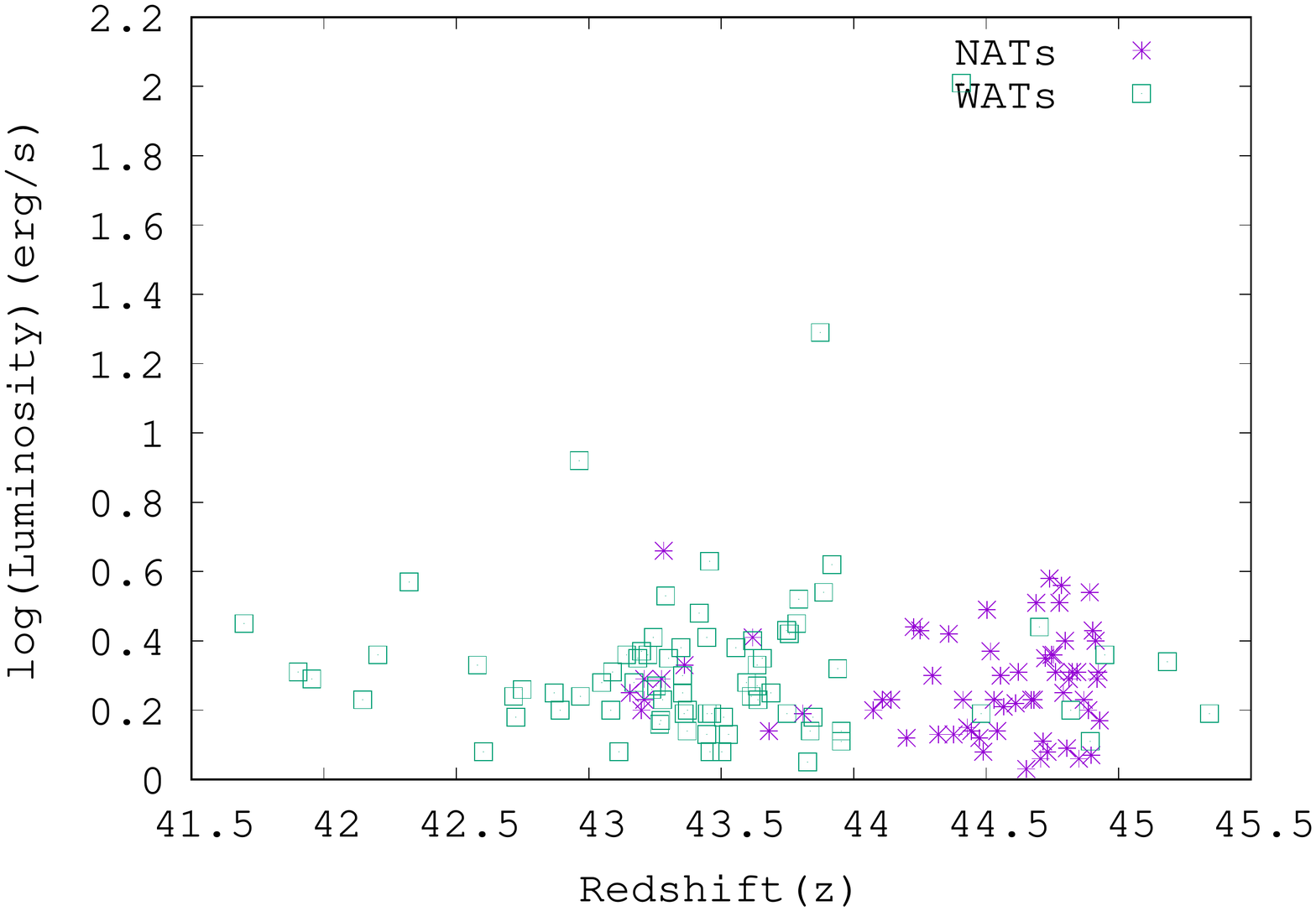}
	\caption{Distribution of radio luminosity ($L_{rad}$) with redshift ($z$).}
    \label{fig:lum-redshift}
\end{figure*}

\begin{figure*}
	\vbox{	
	\centering
\includegraphics[width=8.1cm,origin=c]{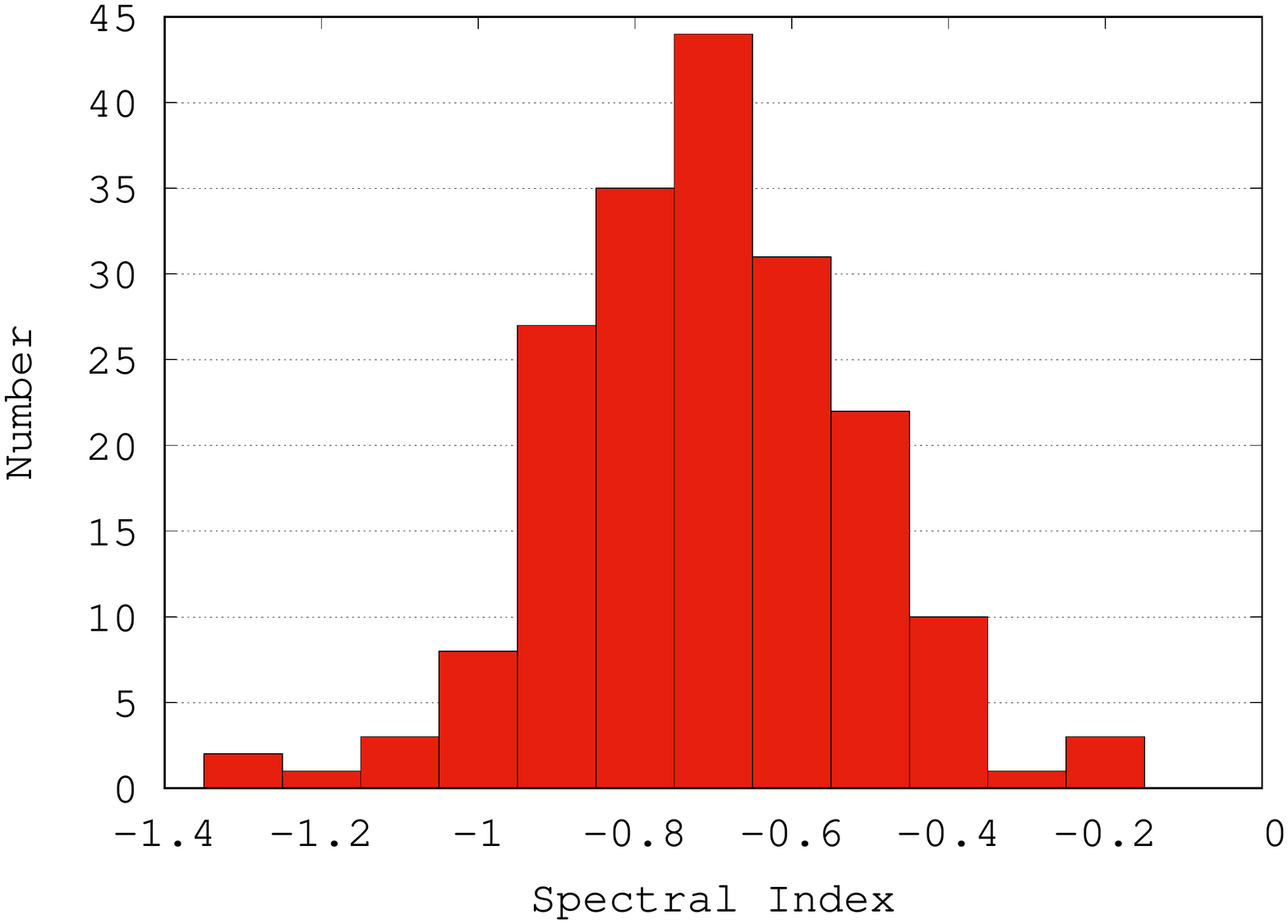}
\includegraphics[width=8.1cm,origin=c]{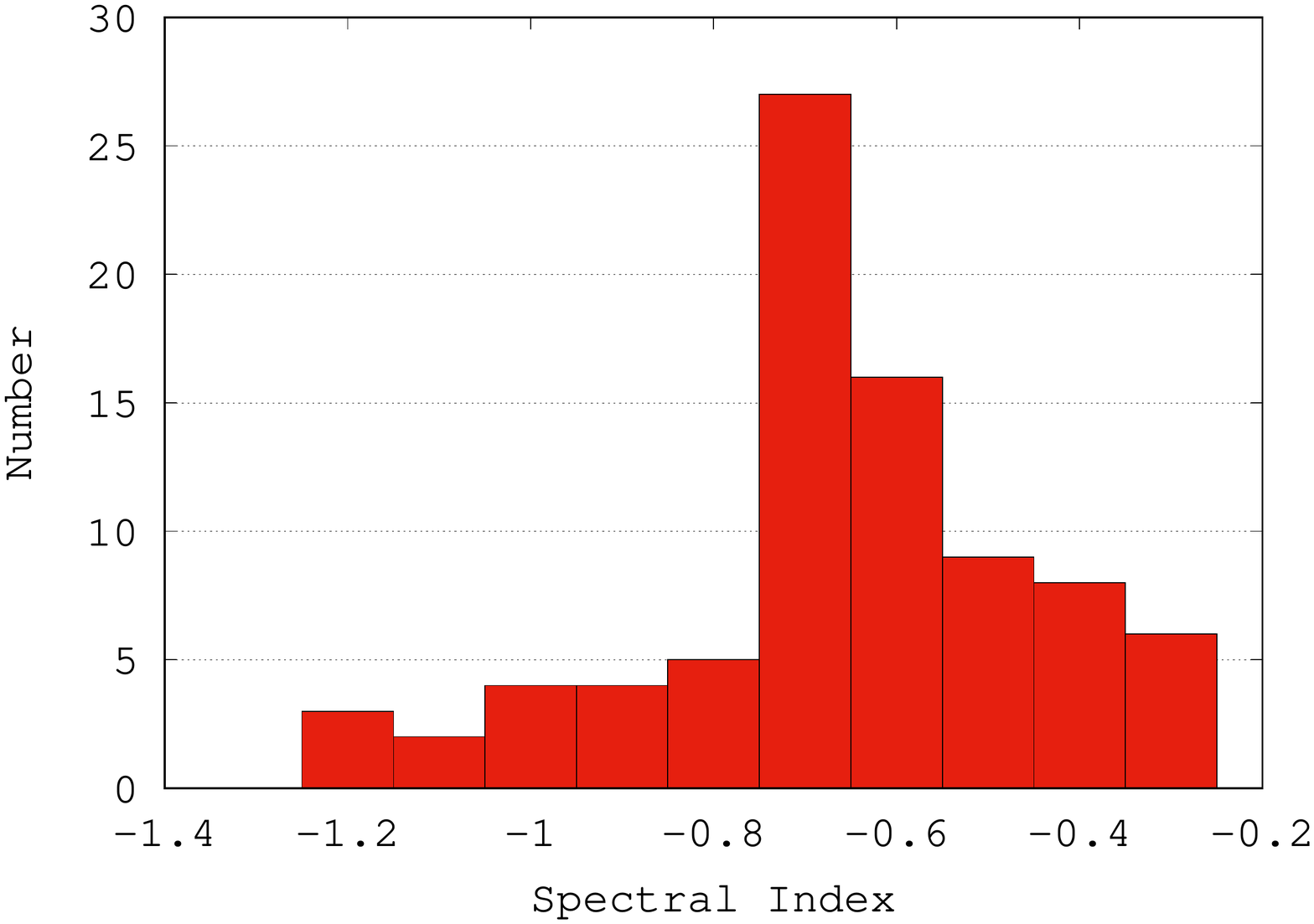}
}
	\caption{Histogram showing spectral index ($\alpha_{1400}^{144}$) distribution of WAT (left) and NAT (right) sources presented in the current paper.}
\label{fig:spindex_histgrm}
\end{figure*}

\begin{figure*}
	\vbox{	
	\centering
\includegraphics[width=8.1cm,origin=c]{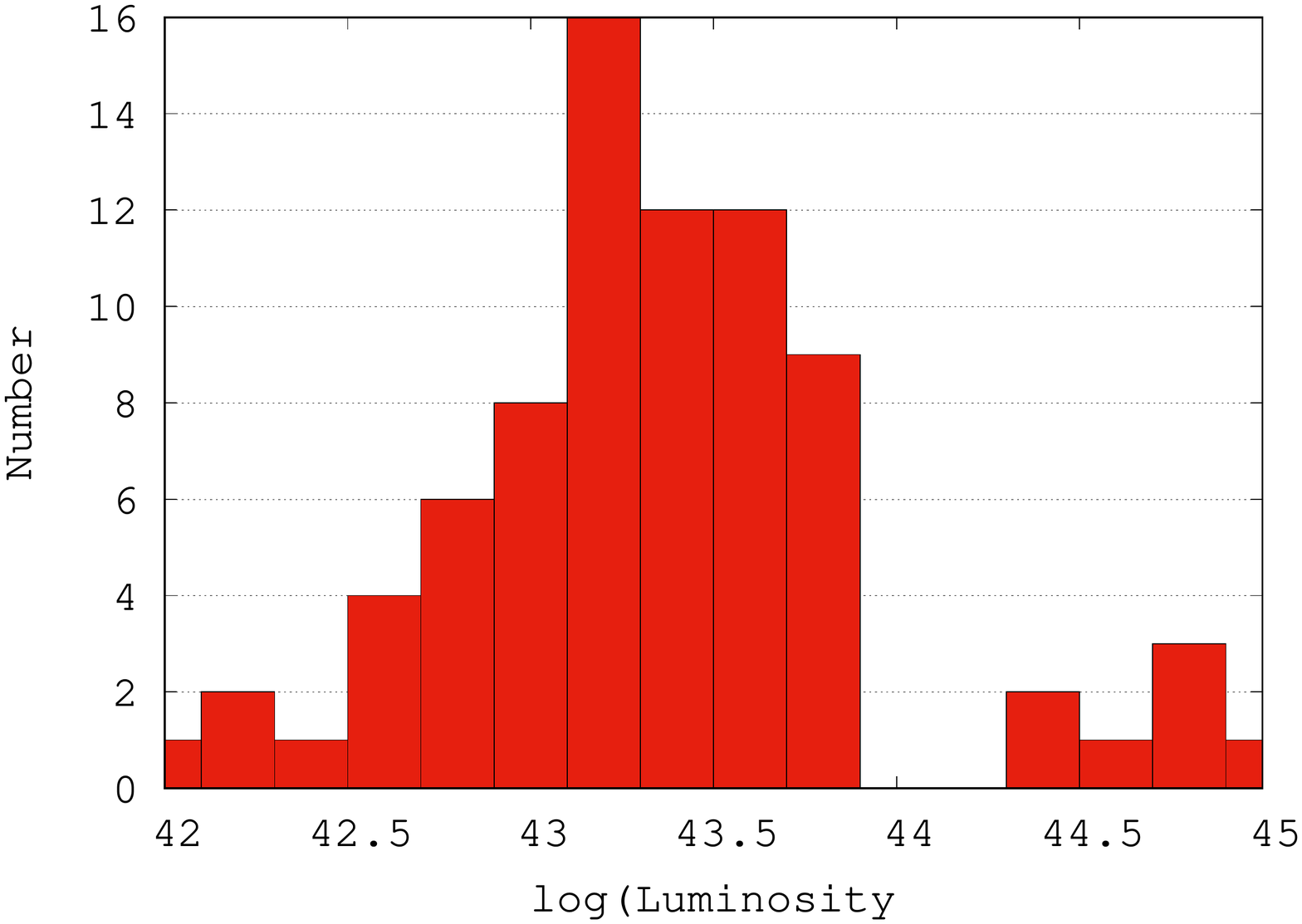}
\includegraphics[width=8.1cm,origin=c]{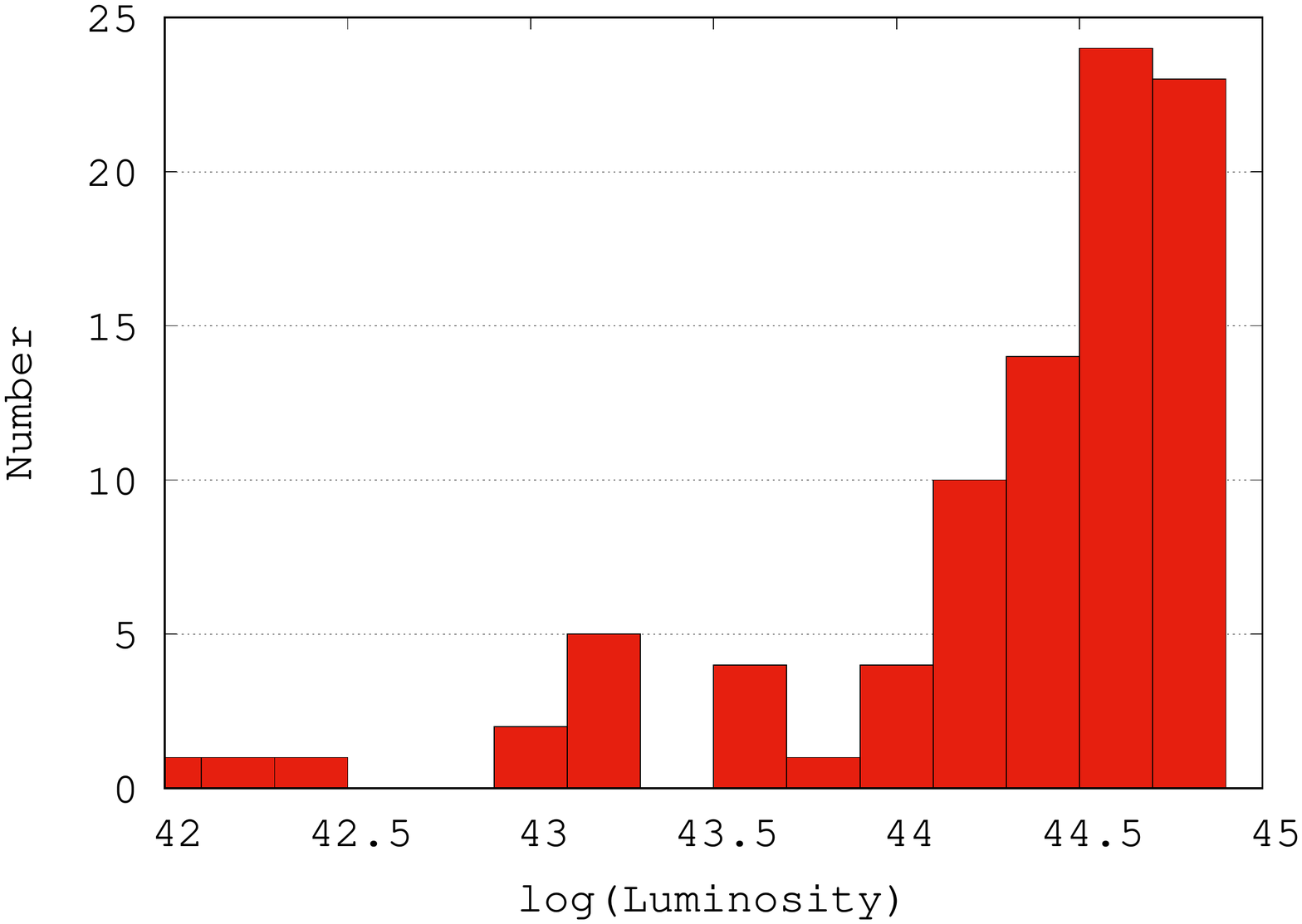}
}
	\caption{Histogram showing distribution of radio luminosity ($L_{1400}$) for WATs (left) and NATs (right).}
   \label{fig:pow-number}
\end{figure*}

\section{Methodology}
\subsection{The FIRST Survey Data}
\label{subsec:First}
The FIRST survey \citep{Be95, Wh96} at 1400 MHz (21cm) covers a radio sky of 10,575 square degrees near the north and south Galactic caps, which is equivalent to 25\% of all the sky. This survey covered RA 7.0h to 17.5h and DEC --8.0 deg to +57.6 deg, totalling 8,444 square degrees in north galactic cap and RA 20.4h to  4.0h, DEC --11.5 deg to +15.4 deg, totalling 2,131 square degrees in the southern sky \citep{Be95}. The FIRST survey uses the Very Large Array (VLA) in its B configuration (VLA B). 3-minute snapshots covering a hexagonal grid using 2$\times$7 3-MHz frequency channels centred at 1365 and 1435 MHz\footnote{http://sundog.stsci.edu/first/description.html} are used to map the entire FIRST sky. The raw data is calibrated and cleaned using an automatic pipeline based on routines from the Astronomical Image Processing System (AIPS)\footnote{http://info.cv.nrao.edu/aips/}. 

The FIRST survey has better sensitivity and resolution compared to the earlier NRAO Very Large Array Sky Survey (NVSS) at the same wavelength. With an angular resolution of 45$^{\prime\prime}$ and an RMS of 0.45 mJy, the NVSS in VLA D configuration covers 82 per cent of the celestial sphere \citep{Co98}. Thus the FIRST data has nine times better resolution compared to the NVSS data.

Taking advantage of good sensitivity (a typical RMS of 0.15 mJy) and resolution (5$^{\prime\prime}$), the FIRST database was used earlier to search for a wide variety of radio galaxies such as winged radio galaxies \citep{Ch07, Be20}, compact steep spectrum sources, core dominated triple sources \citep{Ku02, Ma06}, hybrid morphology radio sources \citep{Go00}, giant radio sources \citep{Ku18} and double-double radio galaxies \citep{Pr11}.

\subsection{Search strategy}
We searched for HT radio sources using the VLA FIRST survey. There is a total of 946,432 radio sources in the FIRST survey. All the sources with a size of at least twice the convolution beam size ($>$10$^{\prime\prime}$) are filtered out from the source catalogue. A total of 95,243 sources are retrieved as a result of our filtering. We inspected fields of all the sources ($>$10$^{\prime\prime}$) to search for new HThead tail radio galaxies. We confidently identified a large number of HT sources. 

From our catalogue, we removed all the HT sources found in the previous study of \citet{Pr11}.


\section{Result}
\subsection{Source catalogue}
We discovered 607 new HT sources using the FIRST survey among which 398 are WAT candidates and 216 are NAT candidates. NVSS flux densities are used for flux density measurements at 1400 MHz as FIRST suffers flux density loss due to lack of short spacing in uv coverage. We calculated two-point spectral indices between 1400 and 150 MHz using flux densities from TGSS and NVSS. We also calculated radio luminosities for all sources with known redshifts. The optical counterparts are searched for all discovered HT sources from the Sloan Digital Sky Survey (SDSS) \citep{Gu06} and NED. We identified optical counterparts for 193 WAT and 104 NAT sources. The position of the optical counterpart is used as the position of the corresponding sources in Table 1. For 205 WAT and 112 NAT sources, no optical counterpart is found, and for these sources, we use our best-guessed eye estimated position of the source using the symmetry and morphology of both of the jets in the radio image.

We estimate the redshifts of our HT candidates using SDSS and NED. Total redshifts were found for 214 out of 607 HT candidates among them 138 redshifts for WATs and 76 are NAT candidates.



\subsection{Radio luminosity ($L_{rad}$)}
\label{subsec:lum}
{Radio luminosities are calculated for all the HT sources, where redshift ($z$) is known. The radio luminosities ($L_{rad}$) of WAT and NAT sources are calculated using
\begin{eqnarray}
L_{rad} = \nonumber 1.2\times10^{27}D^2_{\text{Mpc}}S_0{\nu_0^{-\alpha}}(1+z)^{-(1+\alpha)}          \\
\nonumber	 \times(\nu_u^{(1+\alpha)}-\nu_l^{(1+\alpha)})(1+\alpha)^{-1} ~~~\text{erg s$^{-1}$} \\
\end{eqnarray}
\label{eqn:1}
where $S_0$ is the flux density (Jy) at a given frequency $\nu_0$ (Hz), $D_{\text{Mpc}}$ is luminosity distance to the radio galaxy (Mpc), $z$ is the redshift of the source, $\alpha$ is the spectral index ($S \propto \nu^\alpha$) and $\nu_l$ and $\nu_u$ (Hz) are the lower and upper cut-off frequencies \citep{Od87}.
We assume the lower and upper cutoff frequencies as 10 MHz and 100 GHz respectively. In column 11 of Table 1, the luminosities of sources are shown. 

In Figure \ref{fig:lum-redshift}, the distribution of radio luminosity with known redshift ($z$) is shown for NAT and WAT sources presented in the current paper.
The most luminous NAT source in the present sample is J2348+1157 with $L_{rad}$ = 39.65$\times 10^{43}$erg sec$^{-1}$ and most luminous WAT source is J1446+1402 with $L_{rad}$ = 78.17$\times 10^{43}$ erg sec$^{-1}$. The least luminous NAT source in the present sample is J1353+3305 with $L_{rad}$ = 11.32$\times 10^{41}$erg sec$^{-1}$ and the least luminous WAT source is J1253+0604 with $L_{rad}$ = 1.47$\times 10^{39}$erg sec$^{-1}$.

The mean value of luminosity for all the HT sources presented in the current paper is $\log L$ = 43.91 erg sec$^{-1}$.
For WAT sources, the mean $\log L$ (erg sec$^{-1}$) and median $\log L$ (erg sec$^{-1}$) values of luminosities are 43.40 and 44.60 respectively. 
Similarly, for NAT sources, the mean $\log L$ (erg sec$^{-1}$) and median $\log L$ (erg sec$^{-1}$) values of luminosity are 44.37 and 44.62 respectively. Mean and median luminosities of WAT sources are a little less than that of NAT sources.

\subsection{Spectral Index ($\alpha_{150}^{1400}$)}
The two-point spectral index of newly discovered radio galaxies between 1400 and 150 MHz is calculated assuming $S \propto \nu^{-\alpha}$, where $\alpha$ is the spectral index and $S_{\nu}$ is the radiative flux density at a given frequency $\nu$.
The spectral indexes are mentioned in column 9 of Table 1. In Figure \ref{fig:spindex_histgrm}, the the spectral index ($\alpha_{1400}^{144}$) distribution of 83 WATs and 43 NATs are shown. 
The histogram shows its peak near $\alpha_{150}^{1400}$ = --0.75 for both NATs and WATs sources.
 The histogram shows that the total span of $\alpha_{1400}^{150}$ is from --0.20 to --1.06 for NAT sources. Among all the NAT sources, J0758+4406 have the lowest spectral index with $\alpha_{1400}^{144}$ = --0.20 and J2249+0209 have the highest spectral index with $\alpha_{1400}^{150}$ = --1.06. Similarly total span of ($\alpha_{1400}^{144}$) is from --0.15 to --1.27 for WATs sources. J1457+0232 have the lowest spectral index with $\alpha_{1400}^{150}$ = --0.15 and J1300+2916 have the highest spectral index with $\alpha_{1400}^{150}$ = --1.27. 
The mean and median values of the spectral index of WATs are --0.66 and --0.67 respectively. Similarly for NATs sources mean and median values are --0.61 and --0.61 respectively. That means spectral index values of WATs and NATs are more or less similar to normal-sized radio galaxies. 

\section{Discussion}
\label{sec:5} 
Redshift was found for a total of 35\% (214 out of 607) sources. Seventy (70) NATs sources have redshift value $z$$\leq$ 0.5 and only six have redshift $z\geq$ 0.5. 

The brightest WAT candidate is J1143+5201 with flux density $F_{1400}$=2541 mJy.
High-resolution, more sensitive observations are needed for the confirmation of the nature of these sources. The present study significantly increases the number of known HT sources. In future, more HT sources should be discovered using high sensitive, high-resolution observations.

\section{Conclusion}
We build a new catalogue of a total of 607 HT radio galaxies from the FIRST survey, among them 398 are WATs and 216 are NATs. The present catalogue significantly increases the number of known WATs and NATs. Optical counterparts are found for 297 HT sources. The average value of radio luminosity of WAT and NAT sources is less than that of a typical double-lobed radio galaxy.

\section*{Acknowledgments}
This research has made use of the NASA/IPAC Extragalactic Database (NED) which is operated by the Jet Propulsion Laboratory, California Institute of Technology, under contract with the National Aeronautics and Space Administration. This publication makes use of data products from the Two Micron All Sky Survey, which is a joint project of the University of Massachusetts and the Infrared Processing and Analysis Center/California Institute of Technology, funded by the National Aeronautics and Space Administration and the National Science Foundation. 
}

\end{document}